\documentclass[10pt,twocolumn,letterpaper]{article}

\usepackage{iccv}              

\definecolor{iccvblue}{rgb}{0.21,0.49,0.74}
\usepackage[pagebackref,breaklinks,colorlinks,allcolors=iccvblue]{hyperref}
\usepackage{here}
\usepackage{amsmath}

\def\paperID{*****} 
\def\confName{ICCV}
\def\confYear{2025}

\title{OnomatoGen: Onomatopoeia Generation with the Alpha-Channel in Manga}

\author{Takara Taniguchi\\
The University of Tokyo\\
{\tt\small taniguchi@nlab.ci.i.u-tokyo.ac.jp}
\and
Wataru Shimoda\\
CyberAgent\\
{\tt\small wataru\_shimoda@cyberagent.co.jp}
\and
Kota Yamaguchi\\
CyberAgent\\
{\tt\small yamaguchi\_kota@cyberagent.co.jp}
\and
Hideki Nakayama\\
The University of Tokyo\\
{\tt\small nakayama@ci.i.u-tokyo.ac.jp}
}

\begin{document}
\maketitle
\begin{abstract}
Onomatopoeia is an important element for textual messaging in manga.
Unlike character dialogue in manga, onomatopoeic expressions are visually stylized, with variations in shape, size, and placement that reflect the scene’s intensity and mood.
Despite its important role, onomatopoeia has not received much attention in manga generation.
In this paper, we focus on onomatopoeia generation and propose OnomatoGen, which stylizes plain text to an onomatopoeic style.
We empirically show the unique properties of onomatopoeia generation, which differ from typical text stylization methods, and that OnomatoGen can effectively stylize plain text in an onomatopoeia style.
\end{abstract}    
\section{Introduction}
\label{sec:intro}
Manga is a widely popular form of entertainment that attracts a broad audience, conveying stories through pages of black-and-white illustrations.
A page in manga typically consists of character illustrations, scene illustrations, effects with screen tone, and speech bubbles arranged within panel layouts.
While most textual communication in manga is conveyed through speech bubbles containing character dialogue, there are many other textual elements used to tell stories effectively.
In manga, onomatopoeia is a form of textual expression frequently used for conveying sound effects, emotions, and motion, enhancing the immersive experience. 
Unlike character dialogue, onomatopoeic expressions are visually stylized, with variations in font, size, and placement that reflect the scene's intensity and mood. 
Interestingly, the style and use of onomatopoeia can be distinctive features of a manga's artwork.

In the comic market, the speed of content supply is important, weekly serialized stories are a popular format in Japan, and producing manga on a weekly basis can sometimes lead to artist overwork.
Manga generation methods~\cite{Wu_2025_CVPR_diffsensei,chen2024mangadiffusion} and manga-related computer vision methods~\cite{ Sachdeva_2024_ACCV_tailstelltales,Magi_sachdeva_2024,sachdeva2025magiv3} should be able to alleviate overwork and support faster content supply.
While prior work has addressed panel layouts, comical character generation, and story generation, our focus is on onomatopoeia in manga.
Onomatopoeia generation is an underexplored research area despite its importance as a form of representation.


Onomatopoeia is visually stylized, with variations in shape, size, and placement. Each onomatopoeic word is designed from scratch by the comic artist, and there is no unified rule for their appearance. 
The style of onomatopoeia for each word is often one of a kind.
Furthermore, onomatopoeia in manga can appear in various positions, such as within scene illustrations, on characters, and sometimes even overlapping panel layouts, resulting in complex structures behind the onomatopoeia.
Text stylization research has a long history and has made significant progress. 
However, existing methods~\cite{ NEURIPS2023_chen_textdiffuser,tuo2024anytext, xie2025textfluxocrfreeditmodel,lan2025fluxtextsimpleadvanceddiffusion} are primarily designed for text in scene images, logos, and graphic designs, whereas the properties of onomatopoeic text differ significantly from these of text. 





\begin{figure*}[t]
  \centering
  \includegraphics[width=1.0\textwidth]{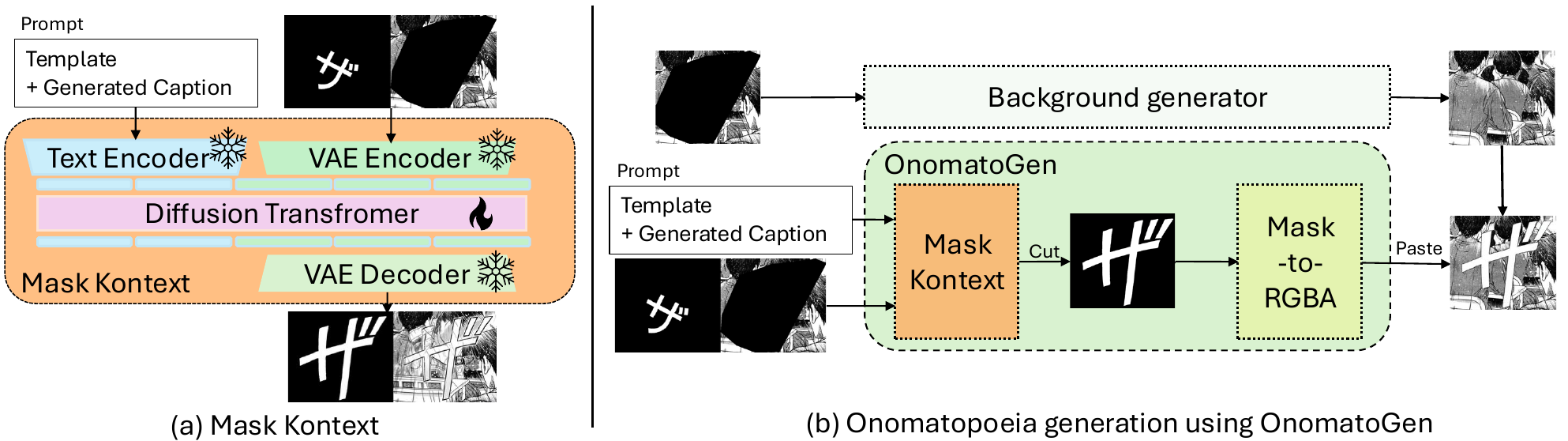}
  \caption{The overall architecture of OnomatoGen. (a) illustrates the architecture of Mask Kontext while (b) shows onomatopoeia generation using OnomatoGen. }\label{fig:workflow}
\end{figure*}

In this paper, we propose OnomatoGen, which stylizes plain text in an onomatopoeic style.
To capture the properties of onomatopoeia, OnomatoGen is designed as a two-step method consisting of stylized mask generation and mask-to-RGBA conversion.
We believe that this modeling enables the handling of complex and unique word shapes, and that enforcing the output in RGBA format helps reduce pixel distortions behind onomatopoeia.
Experimental results demonstrate that onomatopoeia stylization is significantly different from typical text stylization, and OnomatoGen effectively handles the stylization of the onomatopoeia property. 
We also verify that typical conditional visual text generation methods in the RGB format suffer from the simultaneous generation of complex background pixels behind the onomatopoeia. In contrast, OnomatoGen alleviates this issue by separating onomatopoeia text generation in the RGBA format from background generation.

The contributions are summarized as follows:
\begin{itemize}
    \item We propose OnomatoGen, a method designed to stylize text in onomatopoeic form in manga, generating unique shapes while minimizing pixel distortions behind the onomatopoeia.
    \item We empirically demonstrate that onomatopoeia exhibits unique characteristics compared to the texts that typical text stylization methods have focused on, and that OnomatoGen effectively handles these characteristics.
\end{itemize}

\section{Related Work}

Since using generative models in manga has been investigated, recent works addressed the panel-wise Manga generation~\cite{Wu_2025_CVPR_diffsensei,chen2024mangadiffusion}.
In particular, Wu \textit{et al.}~\cite{Wu_2025_CVPR_diffsensei} tackled the panel-wise manga generation.
Additionally, Su \textit{et al.}~\cite{Su_2021_mangagan} enabled style transfer of Manga for face images.
Despite this broad progress, the generation of onomatopoeia in manga remains a largely unexplored area.

Text stylization has a long history.
Patch-based text effect transfer methods~\cite{Awesometypo_2017_CVPR,Yang_contextaware_2018_patch} have been proposed, and GAN-based methods~\cite{Men_2018_CVPR_ganframework,
Yang_TETGAN_2019,Wangintelligenttextstyletransfer_2019_CVPR,Huang2023simplifiedtetgan, 
Yang_2019_ICCV_shapematchingGAN} also actively explored to stylize text effectively. 
Recently, diffusion-based text stylization methods are popular approaches~\cite{WAIsemantictypo2023iluz,Tanveer_2023_ICCV_dsfusion}.
In other literature, visual text generation is also a popular research area, which can draw stylized texts based on specified prompts~\cite{lan2025fluxtextsimpleadvanceddiffusion,NEURIPS2023_chen_textdiffuser,xie2025textfluxocrfreeditmodel,tuo2024anytext}. 
However, the prior work focuses on scene texts, logos, graphic designs, and text stylization in an onomatopoeic style is an unexplored problem. 
OnomatoGen consists of two steps: mask generation and mask-to-RGBA conversion. 
\Cref{fig:workflow} shows the overview of OnomatoGen.
In the mask generation stage, referred to as Mask Kontext, OnomatoGen generates an onomatopoeia shape $x_{m}$ for a plain text rendered image $y_{m}$, a manga context $y$, and a prompt $c$~\footnote{prompt is a combination of a pre-defined template and a generated caption by Qwen-2.5-VL-7B~\cite{bai2025qwen25vltechnicalreport} about the manga context}, where the manga context is an image that includes a polygonal specification indicating where the onomatopoeia should be placed, along with the surrounding pixels.
In the mask-to-RGBA conversion stage, OnomatoGen stylizes the binary mask into an RGBA format using a pre-trained RGBA image generation model~\cite{zhang2024layerdiffuse}. 
Finally, the RGBA onomatopoeia is composited with generated background pixels behind the onomatopoeia.


\subsection{Mask Kontext}\label{subsec:train}
In mask generation, we aim to generate a plausible shape of onomatopoeia from a plain-text rendered image with contextual inputs. We treat this image conversion as a form of image editing and therefore utilize the FLUX.1 Kontext model~\cite{labs2025flux1kontextflowmatching}. 
Our goal is to approximate the conditional distribution $p$ parameterized by $\theta$:
\begin{equation}
 p(x_{m}|y_{m}, c; \theta), 
 \label{eq:fluxkontext}
\end{equation}
which generates $x_{m}$ from the plain text $y_{m}$ and prompt $c$.

To efficiently handle the additional manga context $y$, we extend FLUX.1 Kontext by incorporating In-Context LoRA~\cite{huang2024incontextloradiffusiontransformers}, which we refer to as Mask Kontext.
In-context LoRA can handle visual context by simply concatenating it in the RGB space of the input $x_m$, then we replace the conditional distribution by:
\begin{equation}
 p(\mathrm{Concat}(x_{m}, x) | \mathrm{Concat}(y_{m}, y), c ; \theta),  
 \label{eq:incontextlora}
\end{equation}
where $\textrm{Concat( )}$ means a spatial horizontal concatenation operation and $x$ is an onomatopoeia rendered RGB image. $\textrm{Concat}(x_{m}, x)$ is easily inverted to $x_{m}$ by cutting.
The loss function for training follows FLUX.1 Kontext.

\subsection{Mask-to-RGBA Conversion}
We convert the mask into an RGBA format by LayerDiffuse~\cite{zhang2024layerdiffuse} with the expectation that compositing it with the background will result in fewer pixel distortions behind onomatopoeia. 
First, we sample $\mathrm{Concat}(x_{m},x)$ from the conditional probability from Mask Kontext and obtain $x_{m}$ by spatially separating the concatenated output.
Then, we stylize the mask into an RGBA format $x_{\textrm{rgba}}$ by the pre-trained $\textrm{LayerDiffuse}$:
\begin{equation}
 x_{\textrm{rgba}} \sim \textrm{LayerDiffuse}(x_{m}, c'), 
 \label{eq:layerdiffuse}
\end{equation}
where $c'$ is a pre-defined prompt for converting mask to onomatopoeia style.
Finally, we paste the RGBA onomatopoeia $x_{\textrm{rgba}}$ onto an hole inpainted image $y_{b}$, where $y_{b}$ is inpainted by FLUX.1 Kontext~\cite{labs2025flux1kontextflowmatching} from $y$.

\section{Experiment}\label{sec:experiment}

\begin{figure*}[t]
  \centering  
  \begin{tabular}{ccccccc}
    \toprule 
    \textbf{\shortstack{Plain text \\ w/ LayerDiffuse}} & \textbf{Anytext} & \textbf{TextFlux} & \textbf{FluxText} & \textbf{FLUX.1 Kontext} & \textbf{Ours} & \textbf{GT} \\
    \midrule 
    \includegraphics[width=0.11\textwidth]{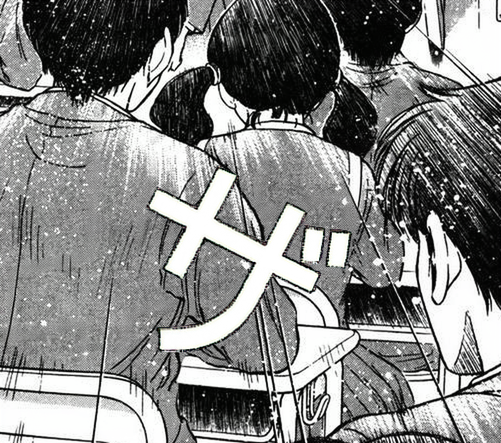} &
    \includegraphics[width=0.11\textwidth]{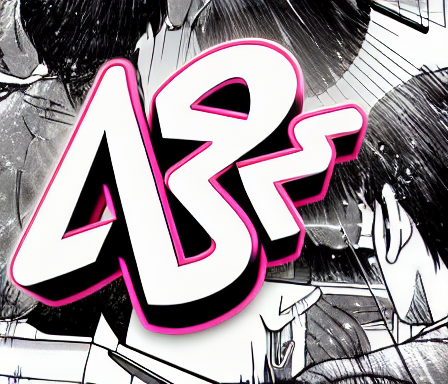} &
    \includegraphics[width=0.11\textwidth]{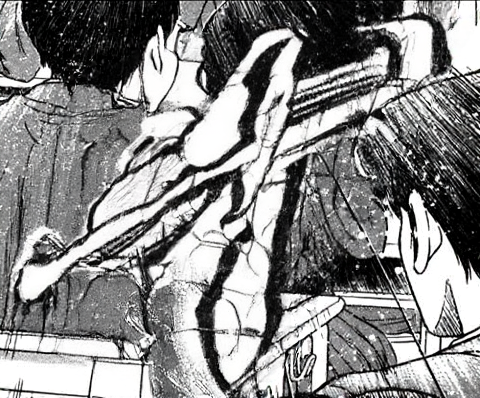} &
    \includegraphics[width=0.11\textwidth]{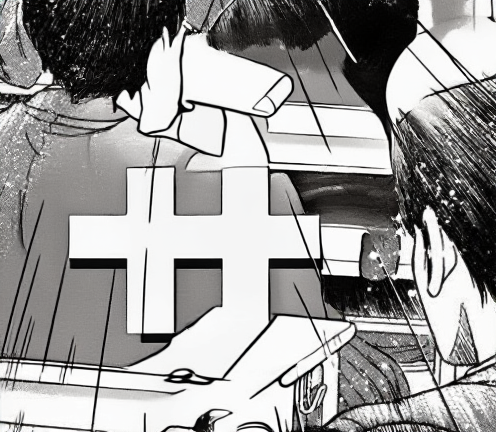} &
    \includegraphics[width=0.11\textwidth]{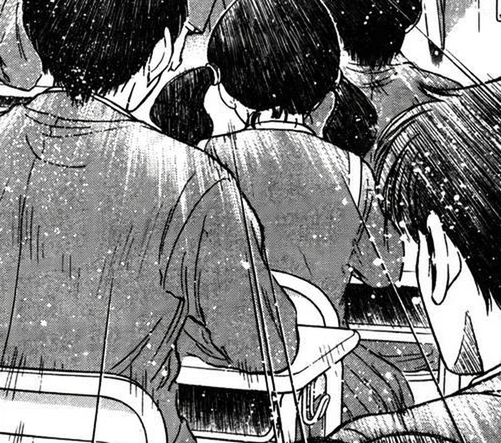} &
    \includegraphics[width=0.11\textwidth]{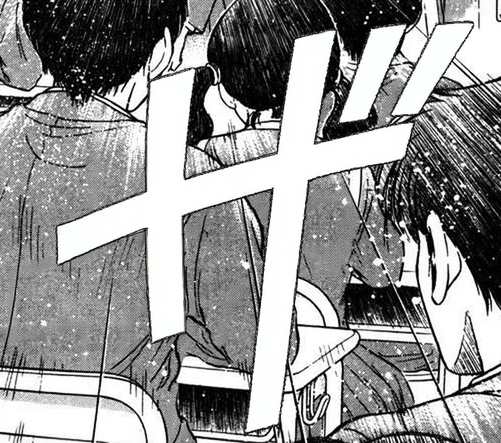} &
    \includegraphics[width=0.11\textwidth]{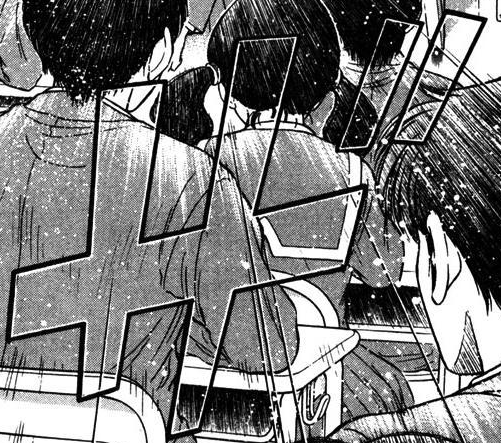} \\
    
    \includegraphics[width=0.11\textwidth]{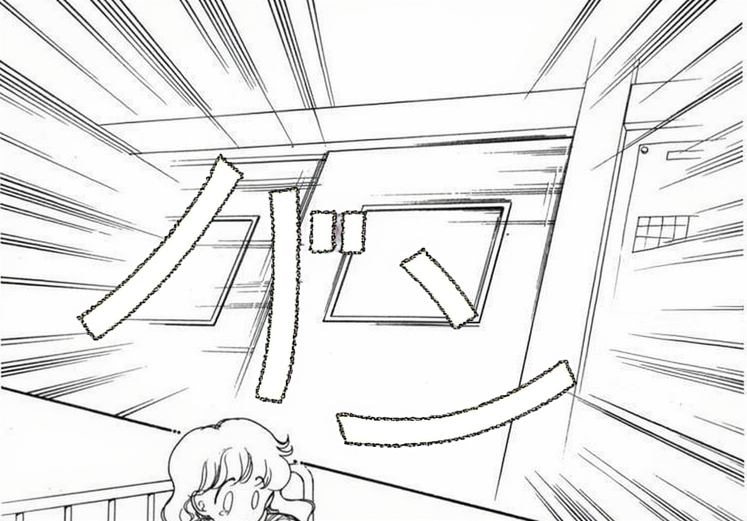} &
    \includegraphics[width=0.11\textwidth]{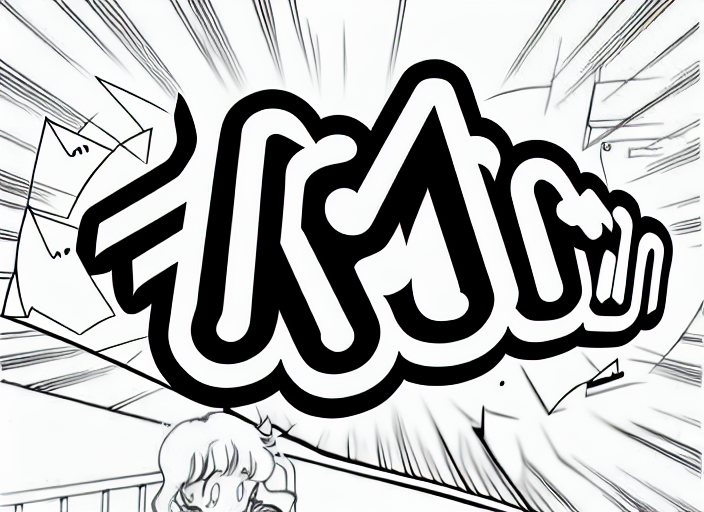} &
    \includegraphics[width=0.11\textwidth]{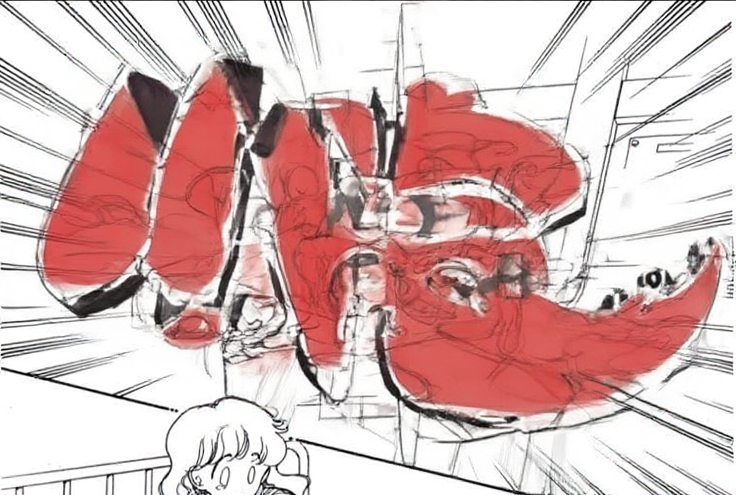} &
    \includegraphics[width=0.11\textwidth]{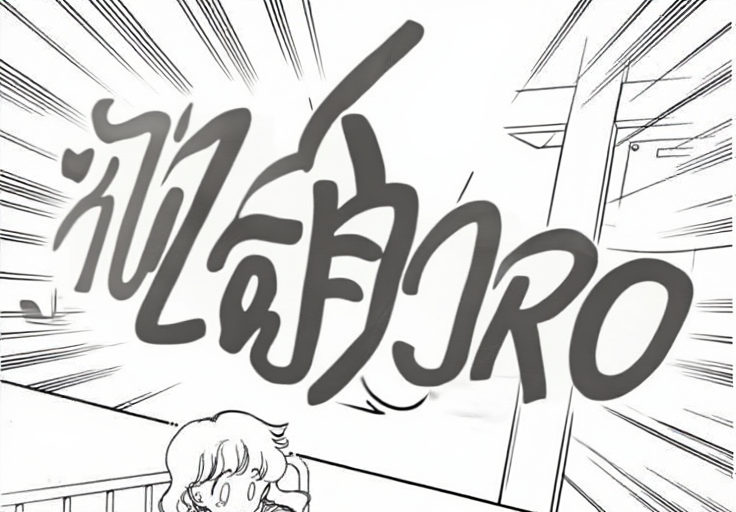} &
    \includegraphics[width=0.11\textwidth]{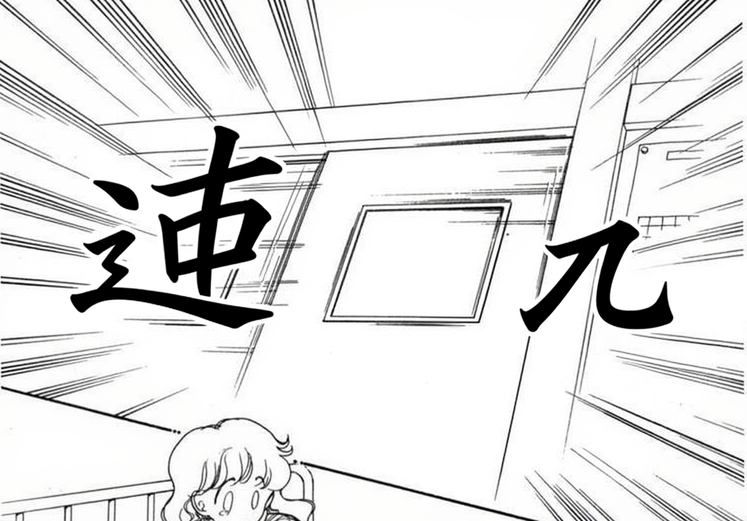} &
    \includegraphics[width=0.11\textwidth]{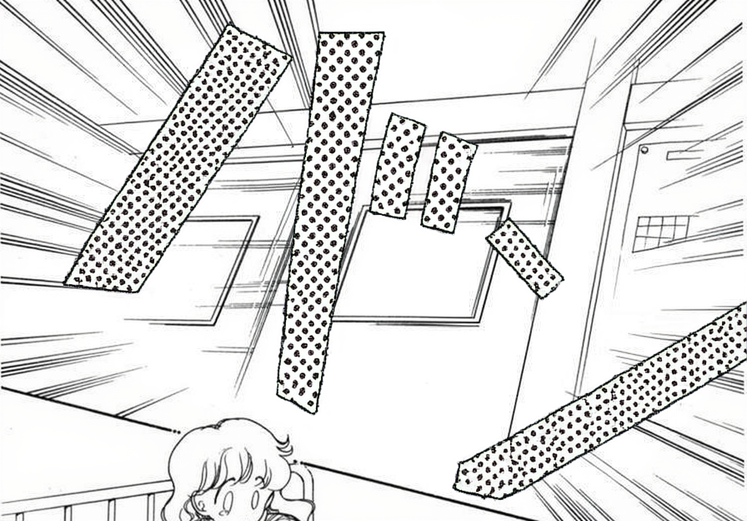} &
    \includegraphics[width=0.11\textwidth]{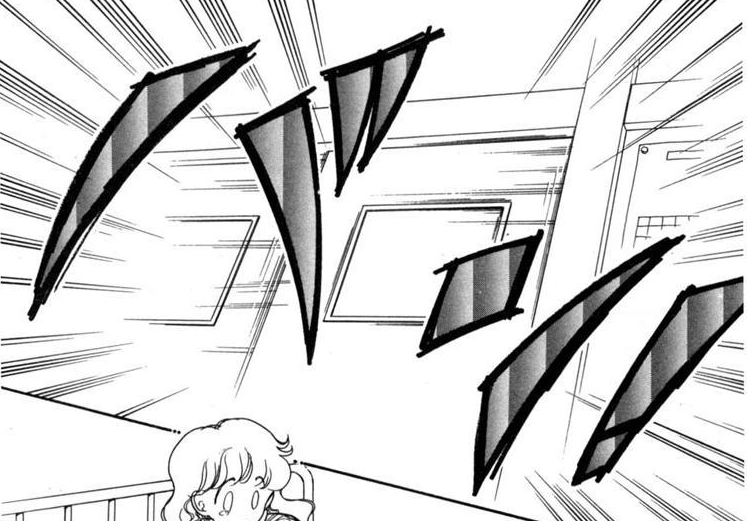} \\
    \bottomrule 
  \end{tabular}
  \caption{Qualitative comparisons of generated onomatopoeia.} 
  \label{fig:comparison_table} 
\end{figure*}

\subsection{Datasets and Evaluation Metrics}
We explain the dataset and evaluation metrics.
We merge the MangaSeg dataset~\cite{Xie_2025_CVPR} and Manga109 onomatopoeia dataset~\cite{baek2022onomatopoeia} as a pair of GT texts and onomatopoeia.
We use images larger than 300$\times$300 pixels because low-resolution images tend to lose critical details of text and onomatopoeia.
Following the split of Manga109~\cite{baek2022onomatopoeia} titles, we used 1,010 images for the train dataset and 169 images for the test dataset, respectively.
We train our method for 25,000 steps using a single NVIDIA A100 GPU with 80 GB of memory. 
To evaluate the performance of our proposed method objectively, we employ the Fr\'echet Inception Distance (FID), and the Normalized Edit Distance (NED) by using the Text Recognition model in COO~\cite{baek2022onomatopoeia}.

\subsection{Comparison Results}
We compare our model with a baseline plain text w/ LayerDiffuse, which simply applies LayerDiffuse~\cite{zhang2024layerdiffuse} to the plain text $y_{m}$, and state-of-the-art methods that can generate stylized texts Anytext~\cite{tuo2024anytext}, TextFlux~\cite{xie2025textfluxocrfreeditmodel}, FluxText~\cite{lan2025fluxtextsimpleadvanceddiffusion}, and FLUX.1 Kontext~\cite{labs2025flux1kontextflowmatching}.
We show the compared results in Table~\ref{tbl:quantitative_results}.
We observe that OnomatoGen outperforms all other competing methods in terms of FID. 
While the baseline shows the highest score on NED due to the lack of textual shape stylization, OnomatoGen achieves a better score on FID while maintaining its NED performance.
\Cref{fig:comparison_table} shows the examples of comparisons.
We observe that state-of-the-art text stylization methods often fail to capture the style of onomatopoeia while preserving text readability.
This result indicates that typical text stylization methods cannot handle onomatopoeic texts, whereas OnomatoGen effectively captures their unique characteristics.




\begin{table}[t]
\centering
\begin{tabular}{@{}lcc@{}}
\toprule
Method      & FID$\downarrow$ & NED$\uparrow$ \\ \midrule
Plain text w/ LayerDiffuse              & 128.6 & 0.258 \\ \midrule
Anytext~\cite{tuo2024anytext}           & 186.8 & 0.070 \\
Textflux~\cite{xie2025textfluxocrfreeditmodel}           & 165.7 & 0.131 \\
Fluxtext~\cite{lan2025fluxtextsimpleadvanceddiffusion}           & 151.0 & 0.092 \\ 
FLUX.1 Kontext~\cite{labs2025flux1kontextflowmatching} & 126.6 & 0.115 \\
OnomatoGen (Ours)               & \textbf{122.7} & \textbf{0.157} \\ \bottomrule
\end{tabular}
\caption{Quantitative comparison of FID and NED.}\label{tbl:quantitative_results}
\end{table}

\section{Ablation Study}
\begin{figure}[t]
  \centering  
  \begin{tabular}{ccc}
    \toprule 
    \textbf{\shortstack{Mask Kontext}} & \textbf{OnomatoGen} & \textbf{GT}  \\ 
    \midrule
    \includegraphics[width=0.11\textwidth]{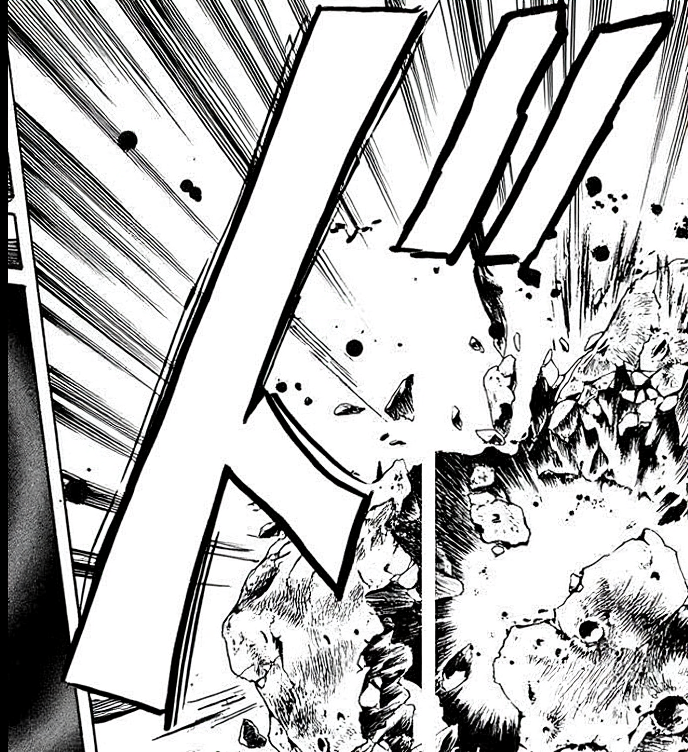} &
    \includegraphics[width=0.11\textwidth]{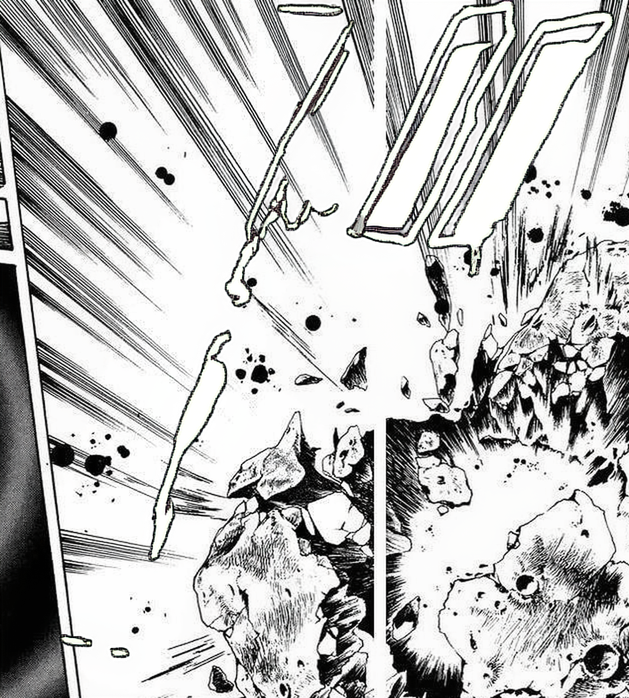} &
    \includegraphics[width=0.11\textwidth]{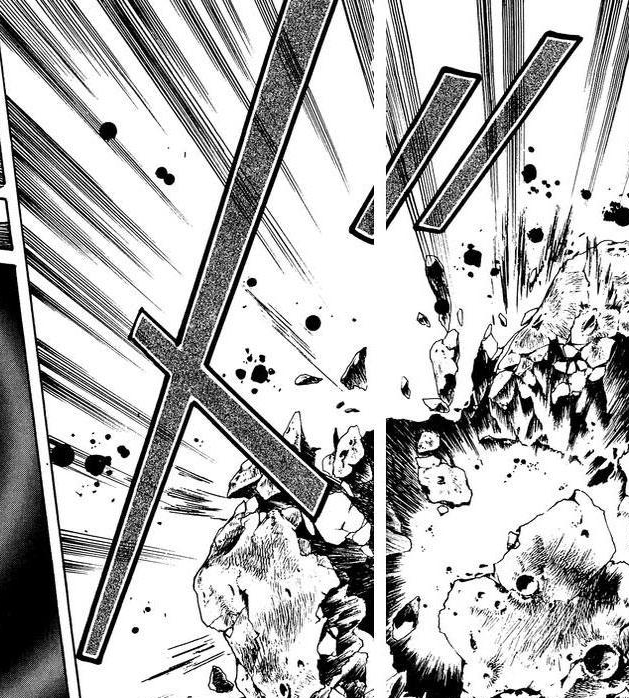} \\
    
    \includegraphics[width=0.11\textwidth]{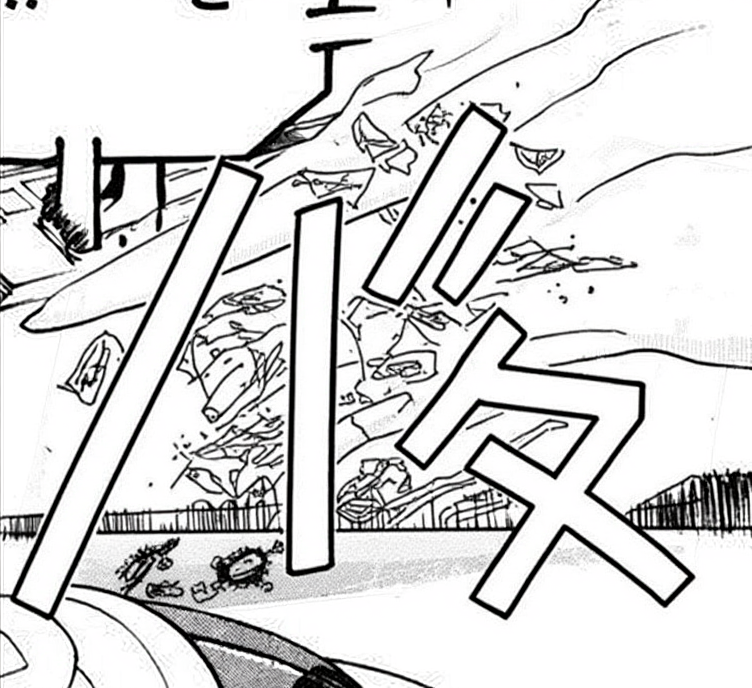} &
    \includegraphics[width=0.11\textwidth]{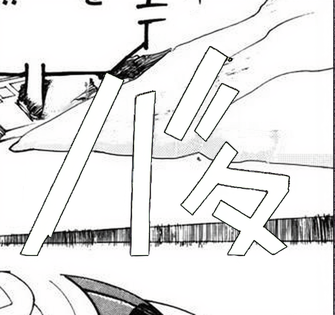} &
    \includegraphics[width=0.11\textwidth]{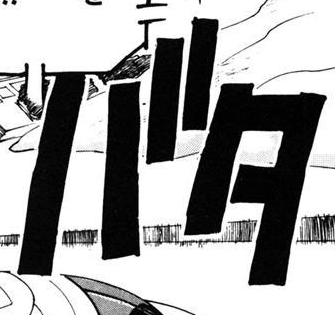} \\
    \bottomrule 
    \end{tabular}
    \caption{Comparisons in Mask Kontext and OnomatoGen.}
    \label{fig:ablation}
\end{figure}

We evaluate the effect of each component of OnomatoGen.


\noindent\textbf{w/o In-Context LoRA} is a fine-tuned FLUX.1 Kontext. 

\noindent\textbf{Mask Kontext} generates an onomatopoeia image by cropping the right half of Mask Kontext's output.

\Cref{tab:ablation_results} shows the effect of the components.
We observe that In-Context LoRA contributes to improvements in both FID and NED scores. In contrast, Mask-to-RGBA conversion improves FID but decreases NED. This is because Mask-to-RGBA conversion can prevent pixel distortions behind the texts, whereas Mask Kontext often fails to generate accurate masks even though it successfully produces the onomatopoeic appearance in RGB pixels.
\Cref{fig:ablation} shows the typical cases. The first row shows that Mask Kontext fails to generate masks but successfully produces the onomatopoeic appearance, resulting in the failure of OnomatoGen. The second row shows Mask Kontext causes pixel distortions behind onomatopoeia, while OnomatoGen generates clear background pixels and onomatopoeia.

\begin{table}[tb]
    \centering
    \begin{tabular}{lcc}
        \toprule
Experiment      & FID$\downarrow$ & NED$\uparrow$ \\ \midrule
w/o In-Context LoRA           & 130.9 & 0.116 \\
Mask Kontext & 146.3 & \textbf{0.277} \\
OnomatoGen & \textbf{122.7} & 0.157 \\ \bottomrule
    \end{tabular}
    \caption{Ablation study for each component of OnomatoGen.}
    \label{tab:ablation_results}
\end{table}

\section{Conclusion}
\label{sec:conclusion}
We propose OnomatoGen, a text stylization method designed for onomatopoeia in manga.
We demonstrate onomatopoeia generation is a significantly different domain from typical text stylization, and that OnomatoGen effectively handles the  properties.

{
    \small
    \bibliographystyle{abbrv}
    \bibliography{main}
}


\end{document}